\documentclass[preprint]{elsarticle}
\usepackage{amssymb}
\usepackage{graphics}
\usepackage{amsmath}
\usepackage{amsfonts}
\usepackage{bm}
\usepackage[]{latexsym}
\usepackage{epsfig}
\usepackage{float}
\usepackage{dsfont}

\newcommand{\be}{\begin{equation}}
\newcommand{\ee}{\end{equation}}
\newcommand{\bea}{\begin{eqnarray}}
\newcommand{\eea}{\end{eqnarray}}
\newcommand{\brr}{\begin{array}}
\newcommand{\err}{\end{array}}
\newcommand{\bit}{\begin{itemize}}\newcommand{\eit}{\end{itemize}}
\newcommand{\ben}{\begin{enumerate}}\newcommand{\een}{\end{enumerate}}

\newcommand{\ba}{\begin{array}}
\newcommand{\ea}{\end{array}}

\def\lf{\left}

\def\pa{\partial}\def\ran{\rangle}

\def\ri{\right}
\def\al{\alpha}
\def\de{\delta}

\def\si{\sigma}
\def\om{\omega}

\def\1{{_{1}}}\def\2{{_{2}}}

\def\noHe0{:\;\!\!\;\!\!:H_e(0):\;\!\!\;\!\!:}
\def\noHm0{:\;\!\!\;\!\!:H_\mu(0):\;\!\!\;\!\!:}

\def\lf{\left}

\def\pa{\partial}\def\ran{\rangle}

\def\ri{\right}

\def\al{\alpha}
\def\de{\delta}

\def\si{\sigma}
\def\om{\omega}

\def\1{{_{1}}}\def\2{{_{2}}}

\begin{document}
\title{Casimir effect for mixed fields}

\author[a,b]{M.~Blasone}
\ead{blasone@sa.infn.it}

\author[a,b]{G.~G.~Luciano}
\ead{gluciano@sa.infn.it}

\author[a,b]{L.~Petruzziello}
\ead{lpetruzziello@na.infn.it}

\author[a,b]{L.~Smaldone}
\ead{lsmaldone@sa.infn.it}

\address[a]{Dipartimento di Fisica, Universit\`a di Salerno, Via Giovanni Paolo II, 132 84084 Fisciano, Italy}
\address[b]{INFN, Gruppo collegato di Salerno, Italy}

  \def\be{\begin{equation}}
\def\ee{\end{equation}}
\def\al{\alpha}
\def\bea{\begin{eqnarray}}
\def\eea{\end{eqnarray}}

  \def\be{\begin{equation}}
\def\ee{\end{equation}}
\def\al{\alpha}
\def\bea{\begin{eqnarray}}
\def\eea{\end{eqnarray}}

\begin{abstract}
We analyze the Casimir effect for a flavor doublet
of mixed scalar fields confined inside a one-dimensional finite region.
In the framework of the unitary inequivalence between mass and flavor
representations in quantum field theory, we employ two
alternative approaches to derive the
Casimir force: in the first case, the zero-point 
energy is evaluated for the vacuum of fields with definite mass, 
then similar calculations are performed for the
vacuum of fields with definite flavor. We find that signatures 
of mixing only appear in the latter context, showing the result to be
independent of the mixing parameters in the former. 
\end{abstract}

 \vskip -1.0 truecm
\maketitle

\section{Introduction}
The concept of vacuum in quantum field theory (QFT) is 
as fascinating as puzzling. In several situations from both 
particle physics and condensed matter, the non-trivial 
condensate structure of the vacuum is crucial 
to explain a variety of both theoretical and observable 
phenomena~\cite{Hawking, Unruh, Bog, Bardeen}. In this connection, 
one of the most eloquent examples is provided by 
the Casimir effect~\cite{casimir}, which occurs whenever 
a quantum field is enclosed in a finite region; 
such a confinement gives rise to a net attractive force between the boundaries, 
the entity of which is closely related to the nature 
of the vacuum itself~\cite{Milton:2001yy}.

In line with these findings, in Refs.~\cite{qftmixing,blascapo} 
it was shown that vacuum also plays a central r\^ole within the 
framework of flavor mixing and oscillations in QFT. 
In Refs.~\cite{qftmixing}, in particular, it was found  
that the vacuum for fields with definite mass (mass vacuum) 
is \emph{unitarily inequivalent}~\cite{Umezawa:1982nv, uir} to the one for fields with 
definite flavor (flavor vacuum), as they are related by a 
non-trivial Bogoliubov transformation. In light of this, it is reasonable
to expect that vacuum effects in the context of QFT mixing 
may, in principle, depend on which 
of these states represents the physical vacuum. 
This is indeed a matter of open
debate~\cite{Giunti:2003dg}:
an interesting 
test bench in this sense has been recently provided by the
analysis of the weak decay of accelerated protons (inverse $\beta$-decay)
with mixed neutrinos~\cite{Ahluwalia:2016wmf,Blasone:2018czm,Cozzella:2018qew}.

Led by these considerations, here 
we analyze the Casimir effect for a system of two  
mixed scalar fields, showing that
the force is sensitive to the choice of the vacuum state. In
particular, we find that the result obtained using the flavor vacuum exhibits corrections that explicitly 
depend on the mixing angle 
and the mass difference of fields, in contrast with the case of the mass vacuum. 

We remark that, although limited to scalar fields in $1+1$ dimensions, our analysis contains all the essential features of the problem, thus giving general validity to our results.
We also stress that the local nature of the Casimir force prevents our calculations from being affected by the choice of a particular regularization scheme. Such a characteristic is not present in other contexts where effects of the flavor vacuum were studied~\cite{Blasone:2004hr}.



The paper is organized as follows: Sec.~\ref{Cefamsc} is devoted to  
briefly review the derivation of the Casimir force for a massive 
scalar field in $1+1$ dimensions. In Sec.~\ref{The problem of inequivalent vacua}, 
we analyze how the standard expression gets modified in the presence of mixed fields by 
performing calculations on both mass and flavor vacua. 
Sec.~\ref{Discussion} contains conclusions 
and an outlook for future developments.
Throughout the paper, we use natural units and the metric in the conventional timelike signature.

\section{Casimir effect for a massive scalar field}
\label{Cefamsc}
Let us start by deriving the Casimir force for a massive 
charged scalar field $\hat{\phi}$ in $1+1$ dimensions 
(to this aim, we basically follow the treatment
of Ref.~\cite{Mobassem}). In this framework, the free 
Lagrangian density $\hat{\mathcal{L}}$ takes the form\footnote{To simplify the 
notation, we shall omit the $(t, x)$-dependence of the field when 
unnecessary.}
\be\label{lag}
\hat{\mathcal{L}}\,=\,\partial_\mu\hat{\phi}^\dagger\partial^\mu\hat{\phi}-m^2\hat{\phi}^\dagger\hat{\phi}\,,
\ee
where $m$ is the mass of the field.

The Dirichlet boundary conditions imposed by the presence of 
the Casimir plates read 
\begin{equation}
\hat{\phi}(t,0)=\hat{\phi}(t,L)=0\,,
\end{equation}
with $L$ being the distance between the two confining surfaces.
These constraints only allow modes with momentum 
$k_n={\pi n}/{L}$ to give a non-vanishing contribution 
to the field expansion, yielding
\be\label{fieldexp}
\hat{\phi}(t,x)\,=\,
\frac{1}{\sqrt{2L}}\sum_{n\,=\,1}^{\infty}\frac{\sin k_nx}{\sqrt{\omega_n}}\left[\hat{a}_n\,e^{-i\omega_nt}+\hat{b}_n^\dagger\,e^{i\omega_nt}\right],
\ee
where $n=1, 2,...$ and $\omega_n\,=\,\sqrt{k_n^2+m^2}$. Here 
$\hat{a}_n$ ($\hat{b}_n^\dagger$) are the usual annihilation 
(creation) operators of a particle (antiparticle) with momentum $k_n$ 
and frequency $\omega_n$. They are assumed to satisfy the 
canonical bosonic algebra
\be
\left[\hat{a}_n, \hat{a}_{n'}^\dagger\right]\,=\,
\left[\hat{b}_n, \hat{b}_{n'}^\dagger\right]\,=\,\delta_{nn'}\,,\quad\, \forall n,n',\,
\ee
with all other commutators vanishing. The vacuum state
is defined by
\begin{equation}
\hat{a}_n|0\rangle\,=\,\hat{b}_n|0\rangle\,=\,0,\quad  \forall n\,.
\label{vacuum}
\end{equation}
In order to compute the Casimir force, let us now evaluate 
the zero-point energy density of the field as
\be
\varepsilon_0\,=\,\langle 0|\hat{T}_{00}|0\rangle\,,
\ee
where $\hat{T}_{\mu\nu}$ is the stress-energy tensor derived 
from the Lagrangian density Eq.~\eqref{lag}~\cite{Itzykson:1980rh}.
A straightforward calculation leads to
\begin{equation}
\varepsilon_0\,=\,\frac{1}{2L}\sum_{n\,=\,1}^\infty\omega_n\,.
\end{equation}
Using the standard definition of Casimir force~\cite{Milton:2001yy,Mukhanov}
\be\label{exprforce}
F_0\,=\,-\frac{\partial}{\partial L}(L\,\varepsilon_0)\,,
\ee
and exploiting a suitable renormalization scheme~\cite{Mobassem}, 
we finally obtain the following finite expression for the net force 
between the plates:
\be\label{finforce}
F\,=\,-\frac{m^2}{\pi}\sum_{n\,=\,1}^\infty\left[K_2(2\hspace{0.2mm}m\hspace{0.2mm}L\hspace{0.2mm}n)-\frac{K_1(2\hspace{0.2mm}m\hspace{0.2mm}L\hspace{0.2mm}n)}{2\hspace{0.2mm}m\hspace{0.2mm}L\hspace{0.2mm}n}\right],
\ee
where $K_{\nu}(x)$ is the modified Bessel function 
of the second kind~\cite{stegun}. Notice that, in the limit $m\rightarrow 0$, 
Eq.~(\ref{finforce}) correctly reproduces the more familiar 
expression of the Casimir force for a massless field~\cite{Milton:2001yy,Mobassem,Mukhanov}.
\section{Casimir effect for mixed fields}
\label{The problem of inequivalent vacua}
Let us now generalize the above formalism to the context 
of field mixing. For this purpose, consider the following 
Lagrangian density describing two charged scalar fields with a
mixed mass term~\cite{blascapo}:
\be\label{lagmix}
\hat{\mathcal{L}}\,=\,\sum_{\si=A,B}\lf(\partial_\mu\hat{\phi}_\si^\dagger\partial^\mu\hat{\phi}_\si-m_\si^2\hat{\phi}_\si^\dagger\hat{\phi}_\si\ri)\,-\,m_{AB}^2\lf(\hat{\phi}^\dagger_A\hat{\phi}_B+\hat{\phi}^\dagger_B\hat{\phi}_A\ri)\,,
\ee
where $\hat{\phi}_\si$ ($\si=A,B$) are the fields with definite flavor $\sigma$. 

It is a trivial matter to check that the mixing transformations
\begin{eqnarray}\label{mix1}
\begin{pmatrix}\hat{\phi}_A\\\hat{\phi}_B\end{pmatrix}=\begin{pmatrix}\cos\theta&\sin\theta\\-\sin\theta&\cos\theta\end{pmatrix}\begin{pmatrix}\hat{\phi}_1\\\hat{\phi}_2\end{pmatrix}\,,
\end{eqnarray}
allow to recast the quadratic form Eq.~(\ref{lagmix}) into 
a diagonal Lagrangian density for two free charged scalar
 fields $\hat{\phi}_j$ ($j=1,2$) with mass $m_j$:
\be\label{lag12}
\hat{\mathcal{L}}\,=\,\sum_{j=1,2}\lf(\partial_\mu\hat{\phi}_j^\dagger\partial^\mu\hat{\phi}_j-m_j^2\hat{\phi}_j^\dagger\hat{\phi}_j\ri),
\ee
where the two sets of mass parameters $m_\si$ and $m_j$  
are related by
\begin{eqnarray}
\label{masses}
m_A^2&=&  \cos^2\theta\,m_1^2 \,+\, \sin^2\theta\,m_2^2\,,\\[2mm]
m_B^2&=&  \sin^2\theta\,m_1^2 \,+\, \cos^2\theta\,m_2^2\,,
\label{masses2}
\end{eqnarray}
and $m^2_{AB}$ in Eq.~(\ref{lagmix}) is given 
by $m^2_{AB}=(m_2^2-m_1^2)\,\sin\theta\cos\theta$. 
The mixing angle $\theta$ is defined 
as $\tan2\theta={2m_{AB}^2}/{\lf(m_B^2-m_A^2\ri)}$.

Note that each of the two fields 
$\hat\phi_j$ ($j=1,2$) in Eq.~(\ref{mix1}) 
can be expanded as in Eq.~(\ref{fieldexp}). 
Thus, according to Eq.~\eqref{vacuum}, one can 
define the vacuum for fields with definite mass 
(\emph{mass vacuum}) as
\be\label{massvac}
\hat{a}_{n,j}|0\rangle_{1,2}=\hat{b}_{n,j}|0\rangle_{1,2}=0\,, \qquad \forall n, \quad j=1,2\,.
\ee

To derive the corresponding relation for fields with definite 
flavor, it is worth rewriting Eq.~\eqref{mix1} in terms of the mixing 
generator $K_{\theta,\mu}(t)$~\cite{fuji} as:
\bea
\label{Bogofield}
\qquad\hat{\phi}_\chi(t,x) \ = \ K^{-1}_{\theta,\mu}(t) \, \hat{\phi}_l(t,x) \, K_{\theta,\mu}(t) \,, \qquad (\chi,l) \, =\, (A,1), (B,2) \, , 
\eea
where $K_{\theta,\mu}(t)=G_\theta(t)\,I_\mu(t)$, with
\begin{eqnarray}
\nonumber
G_\theta(t)&=&\exp\bigg[-i \theta \int^L_0 \!\! \mathrm{d} x \,\lf(\hat{\pi}_1(t,x)\hat{\phi}_2(t,x)\,+\,\hat{\phi}^\dag_2(t,x)\hat{\pi}^\dag_1(t,x)\ri.\\[1mm]
&&\left.-\,\hat{\pi}_2(t,x)\hat{\phi}_1(t,x)\,-\,\hat{\phi}^\dag_1(t,x)\hat{\pi}^\dag_2(t,x) \ri)\bigg] \,,
\end{eqnarray}
and
\be\label{I}
I_\mu(t) \ = \ \exp\lf[\sum^\infty_{n=1}\sum_{\si,j}\xi^n_{\si,j}\Big(a^\dag_{n,\si}(t)b^\dag_{n,\si}(t)\,-\,b_{n,\si}(t)a_{n,\si}(t)\Big)\ri] \, .
\ee
%
%
%
Here $\hat\pi_j\equiv\partial_t\hspace{0.2mm}\phi_j^\dagger$ ($j=1,2$) 
is the canonical momentum conjugate to the field
$\hat\phi_j$, $\xi^n_{\si,j}\equiv\frac{1}{2}\log\lf(\frac{\om_{n,\si}}{\om_{n,j}}\ri)$ and
$\omega_{n,\si}=\sqrt{k_n^2+\mu_\si^2}$ ($\sigma=A,B$).

For $\mu_A=m_1$ and $\mu_B=m_2$, one can 
easily check that $I_\mu(t)=\mathds{1}$, and the field expansions
for definite flavor fields read
\be\label{newflavor}
\hat{\phi}_\chi(t,x)\,=\,\frac{1}{\sqrt{2L}}\sum_{n\,=\,1}^{\infty}\frac{\sin k_nx}{\sqrt{\omega_{n,l}}}\left[\hat{{a}}_{n,\chi}(t)\,e^{-i\omega_{n,l}t}\,+\,\hat{{b}}_{n,\chi}^\dagger(t)\,e^{i\omega_{n,l}t}\right],
\ee
with $(\chi,l)=(A,1), (B,2)$. The corresponding vacuum (\emph{flavor vacuum}) is defined by
\be\label{fs}
|{0}(t)\ran_{A,B} \ = \ G^{-1}_{\theta} (t) |0\ran_{1,2} \, ,
\ee
with $\hat{{a}}_{n,\si}(t)|{0}(t)\ran_{A,B}\,=\,\hat{{b}}_{n,\si}(t)|{0}(t)\ran_{A,B}\,=\,0\,$, $\forall n, t$, 
as expected\footnote{In the following, 
we will work in the Heisenberg picture: this is particularly 
convenient in the present context since special
care has to be taken with the time dependence of flavor vacuum
(see the discussion in Ref.~\cite{Blasone:1998hf}).}.

We stress that the action of the mixing generator $K_{\theta,\mu}(t)$ 
on the mass vacuum is non-trivial: Eq.~(\ref{Bogofield}), 
indeed, hides a Bogoliubov transformation at the level of ladder operators,
which induces a rich condensate structure into the flavor vacuum.
The crucial point is that, in the infinite volume limit, mass and flavor 
vacua become orthogonal to each other, thus
giving rise to \emph{unitarily inequivalent} Fock spaces~\cite{qftmixing} 
(this is a well-known feature of QFT~\cite{Umezawa:1982nv,uir}, 
reflecting in the non-unitary nature of
the generator of Bogoliubov transformations in the infinite volume limit).

Note that the expansions Eq.~\eqref{newflavor}
rely on a particular choice of the wave function basis, namely that referring to the 
free field masses $m_1$, $m_2$. However, a natural alternative put forward 
in Ref.~\cite{Blasone:2011zz} would be to
expand flavor fields in the basis corresponding to 
$\mu_A=m_A$ and $\mu_B=m_B$, with $m_A$ and 
$m_B$ given in Eqs.~\eqref{masses} and \eqref{masses2},
respectively\footnote{
As shown in Ref.~\cite{Blasone:2011zz}, this setting is singled out by 
the requirement that flavor states must be simultaneous eigenstates of the 4-momentum 
and flavor charge operators.}.
For the sake of completeness, 
in what follows we shall deal with both these two cases;
consistently with the previous notation,
the vacuum associated to the last
expansion will be referred to as \emph{tilde flavor vacuum} and denoted by
$|\widetilde{0}(t)\rangle_{A,B}$.


On the basis of the above discussion, it seems obvious that 
different candidates for the r\^ole of fundamental vacuum must be taken into
account in the context of field mixing: the mass vacuum $|0\rangle_{1,2}$,
the flavor vacuum $|0(t)\rangle_{A,B}$ 
and the tilde flavor vacuum $|\tilde0(t)\rangle_{A,B}$. 
The question naturally arises as to which of these states has
indeed physical meaning: to this end, in the following
we evaluate the Casimir force in the aforementioned cases,
showing that the result carries footprints of the  particular choice of vacuum. 


\subsection{Mass vacuum}
\label{vacuummass}
To begin with, we investigate the Casimir effect by 
assuming the mass vacuum $|0\rangle_{1,2}$ 
to be physical. In this case, it is a trivial matter to check
that calculations closely follow the ones of 
Sec.~\ref{Cefamsc}, giving the
following expression for the Casimir force:
\be\label{finforcemass}
F_m\,=\,-\sum_{j=1,2}\frac{m^2_j}{\pi}\sum_{n\,=\,1}^\infty\left[K_2(2\hspace{0.2mm}m_j\hspace{0.2mm}L\hspace{0.2mm}n)\,-\,\frac{K_1(2\hspace{0.2mm}m_j\hspace{0.2mm}L\hspace{0.2mm}n)}{2\hspace{0.2mm}m_j\hspace{0.2mm}L\hspace{0.2mm}n}\right] \, ,
\ee
where the subscript $m$ is a reminder that we are 
dealing with the mass vacuum. By comparison with
 Eq.~\eqref{finforce}, we find out that the result is 
 the same we would obtain for two non-interacting  
 (unmixed) fields. In other words, the Casimir force 
 evaluated with respect to the vacuum $|0\rangle_{1,2}$ 
 is \emph{insensitive} to the mixing, being independent 
 of the mixing angle $\theta$ (this could be somehow 
 expected, since by definition
the mass vacuum can be factorized into the product of
the vacuum for the field $\hat\phi_1$ times the vacuum
for the field $\hat\phi_2$).

\subsection{Tilde flavor vacuum}
Let us now analyze how the Casimir force 
gets modified when referring to the tilde flavor 
vacuum $|\widetilde{0}(t)\rangle_{A,B}$. 
Following the same line of reasoning of 
Sec.~\ref{vacuummass} and exploiting 
Eqs.~\eqref{newflavor} and \eqref{fs} with $\mu_A=m_A$, 
$\mu_B=m_B$, we have
\be\label{finforceflav}
\widetilde{F}_f\,=\,-\sum_{\si=A,B}\frac{m^2_\si}{\pi}\sum_{n\,=\,1}^\infty\left[K_2(2\hspace{0.2mm}m_\si\hspace{0.2mm}L\hspace{0.2mm}n)-\frac{K_1(2\hspace{0.2mm}m_\si\hspace{0.2mm}L\hspace{0.2mm}n)}{2\hspace{0.2mm}m_\si\hspace{0.2mm}L\hspace{0.2mm}n}\right] \,,
\ee
where the subscript $f$ stands for flavor. 

\begin{figure}[t]
\centering
\includegraphics[scale=1.15]{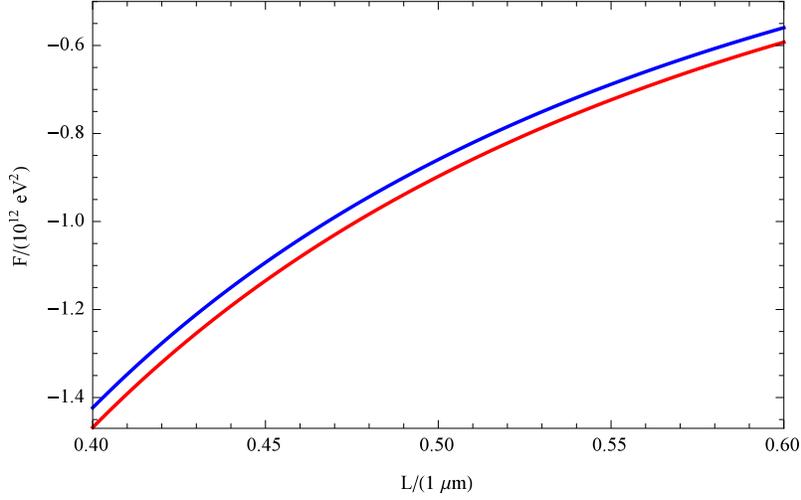}
\caption{The behavior of $F_m$ (blue line) and $\widetilde{F}_f$ 
(red line) as functions of the distance $L$, for sample values 
of $\theta$, $m_1$ and $m_2$ ($\theta=\pi/3$, $m_1=0.8\,\mathrm{eV}$, 
$m_2=0.9\,\mathrm{eV}$).}
\label{fig1}
\end{figure}

The forces $F_m$ and $\widetilde{F}_f$ have been numerically 
evaluated and plotted in Fig.~\ref{fig1} as functions of the 
distance $L$ between the plates. 
We can readily see that their overall behaviors do not differ 
significantly from each other. To enhance the validity 
of this result, we can also compare their analytic expressions, 
at least in the approximation of $\delta m^2L^2\ll 1$, where 
$\de m^2\,\equiv \, m_2^2-m_1^2$. In this regard, 
let us recast Eqs.~\eqref{masses} and \eqref{masses2} in the form
\bea
m_A^2&=&m_1^2 \, + \, \sin^2 \theta \, \de m^2 \, , \\[2mm] 
m_B^2&=&m_2^2 \, - \, \sin^2 \theta \, \de m^2 \,.
\eea
For $\delta m^2L^2\ll 1$, we are allowed to expand 
Eq.~\eqref{finforceflav}  as follows
\be\label{flavorf}
\widetilde{F}_f\,\approx\, F_m \, + \, \de F_1 \, , 
\ee
where the first-order correction $\de F_1$ can be obtained 
by using the asymptotic 
form of the modified Bessel function~\cite{stegun} and 
the zeta function regularization~\cite{Mukhanov}. 
By explicit calculation, we have
\be\label{findelta}
\delta F_1\,\approx\,-\frac{\pi}{12}\frac{\sin^2\theta}{L^2}\frac{\lf(\delta m^2\ri)^2}{m^2_1m^2_2}\,.
\ee
By comparing with Eq.~\eqref{finforcemass}, 
we see that the magnitude of the force
derived with respect to the tilde flavor vacuum is higher
than the one calculated on the mass vacuum (of course,
the gap between $\widetilde{F}_f$ and $F_m$ narrows
as $L$ increases, since $\delta F_1\rightarrow0$ for $L\rightarrow\infty$,
as expected). Remarkably, $\widetilde{F}_f$ explicitly depends on the 
mixing parameters $\theta$ and $\de m^2$, a feature which is absent
in the previous framework. 

\subsection{Flavor vacuum}
Finally, let us evaluate the Casimir force on the 
flavor vacuum $|0(t)\rangle_{A,B}$. In this case, 
using Eqs.~\eqref{newflavor} and \eqref{fs} with $\mu_A=m_1$, 
$\mu_B=m_2$, the zero-point energy density is given by
\be
\varepsilon_0 \, \equiv\, \frac{1}{2L} \sum_{n\,=\,1}^\infty\, \sum_{j=1,2} \, \omega_{n,j}\, \lf(1\,+\,2\hspace{0.2mm}|V_n|^2 \hspace{0.3mm} \sin^2 \theta\ri) \, ,
\ee
where
\be
|V_n| \,\equiv\, \frac{1}{2} \lf(\sqrt{\frac{\om_{n,1}}{\om_{n,2}}}\,-\,\sqrt{\frac{\om_{n,2}}{\om_{n,1}}}\ri) \, .
\ee
Thus, inserting the previous expression into Eq.~\eqref{exprforce}, the Casimir force $F_f$ 
can be cast in the form
\be \label{F1}
F_f \, = \, F_m \ + \ \de F_2 \, , 
\ee
where $F_m$ has been defined  in Eq.~\eqref{finforcemass} and
\be
\label{deltaf1}
\de F_2 \ = \ -\frac{\pa }{\pa L} \lf[ \, \sum_{n\,=\,1}^\infty \lf(\omega_{n,1}
+\omega_{n,2}\ri)\, |V_n|^2 \, \sin^2 \theta \ri]\, .
\ee

\begin{figure}[h]
\centering
\includegraphics[scale=1.2]{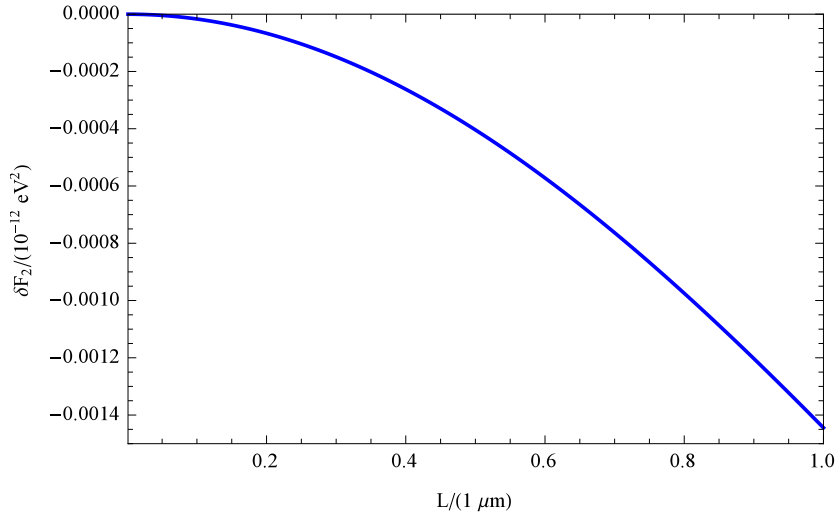}
\caption{The behavior of $\de F_2$ as function of $L$ 
(in the regime of small $L$), for $\theta=\pi/3$, 
$m_1=0.8\,\mathrm{eV}$ and $m_2=0.9\,\mathrm{eV}$.}
\label{fig2}
\end{figure}

The correction $\de F_2$ has been numerically evaluated and
plotted  for small values of $L$ (see Fig.~\ref{fig2}). 

By comparison with Eq.~(\ref{findelta}),
we also provide its analytic expression 
in the limit of $m_j L\ll1\,$ $(j=1,2)$
\be \label{deF1}
\de F_2 \ = \ - \frac{3 \, \sin^2\theta \, L^2 \, \zeta(3) \, \lf(\de m^2 \ri)^2}{2 \pi^3} \, , 
\ee
where $\zeta(3)$ is the Ap\'ery's constant~\cite{apery}. Note that,
within such an approximation, the magnitude of $\de F_2$ grows 
as $L$ increases, as it is evident from Fig.~\ref{fig2}. Clearly, this 
is a subdominant contribution to the total Casimir force $F_f$, 
which indeed decreases as $L$ grows, as it can be easily checked.
On the other hand, the behavior of $\de F_2$ drastically 
changes for large distances, where the approximation 
$m_j L\ll1\,$ fails, and indeed $\de F_2$ correctly vanishes.

Therefore, by comparing Eqs.~\eqref{finforcemass},~\eqref{flavorf} and~\eqref{F1},
we realize that the predicted value of the Casimir force for mixed fields 
varies depending on 
which of the three aforementioned vacua -- the mass, the flavor
and the tilde flavor vacua -- is indeed the physical one.

\section{Discussion and Conclusions}
\label{Discussion}
In this paper, we have addressed the Casimir effect for
a doublet of mixed scalar fields
confined inside a one-dimensional finite region. 
In the framework of the long-standing discussion on the
unitary inequivalence between mass and flavor representations for mixed fields,
we have analyzed to what extent different choices of the
vacuum lead to different predictions for the Casimir force.
Specifically, we have found that, whilst the force computed
with respect to the mass vacuum $|0\rangle_{1,2}$ does not exhibit 
mixing signatures, the use of either the flavor vacuum $|0(t)\rangle_{A,B}$ or
the tilde flavor vacuum $|\widetilde{0}(t)\rangle_{A,B}$ 
shows off an explicit dependence on the characteristic mixing parameters $\theta$
and $\de m^2$. Underpinned by 
future experimental results, the arising discrepancy may provide us 
with the possibility to 
discriminate which of these states
does indeed represent the physical vacuum for mixed fields.


Besides its considerable formal interest,  the issue of inequivalent vacua 
in the context of flavor mixing in QFT is intimately related with 
a series of controversial problems that are currently being put forward in literature.
Recently, for instance, the analysis of the inverse $\beta$-decay with mixed
neutrinos has been fiercely taken back in the spotlight~\cite{Blasone:2018czm,Cozzella:2018qew}
after  a possible disagreement between the decay rates
in the laboratory and accelerated frames was highlighted~\cite{Ahluwalia:2016wmf}.
As pointed out in Refs.~\cite{Blasone:2018czm}, the origin of the contention
is rooted in the choice of asymptotic neutrino states (and, thus, of
vacuum) as mass or flavor eigenstates. Divergent views have been
expressed on the subject~\cite{Blasone:2018czm,Cozzella:2018qew}, and
precious insights toward the ultimate answer 
may be indeed provided by results presented here.

As partially highlighted in the above consideration, the context of accelerated frames is very promising, thus being potentially useful in several other ways. For instance, from the analysis contained in Ref.~\cite{buono}, it is possible to come up with another interesting application for the formalism developed in the present work. Indeed, in Ref.~\cite{buono} it was found that the Unruh radiation acquires a non-thermal contribution in the case of mixed fields. Applied to the Casimir effect, this occurrence can be employed to quantitatively verify whether and how the aforementioned non-thermal corrections affect the mean vacuum energy density and -- as a consequence -- the pressure between the plates. 

Further, we want to point out that flavor vacua can be regarded as \emph{time crystals}~\cite{timecrystal}. In fact, they exhibit a time dependence which is also reflected on physical quantities, such as the fluctuations of the flavor charge.

Finally, it should be remarked that (extended to arbitrary spin fields) our formalism may be exploited 
to fix more stringent constraints on the axion-photon 
and axion-nucleon coupling constants, as well as on the range of 
the axion mass~\cite{Bezerra:2014ona}  
in a context different than the one considered so far~\cite{Gavrilyuk:2018xvk}.
Work in this direction is currently under active investigation~\cite{prog}.


\section*{Acknowledgments}
It is a pleasure to acknowledge helpful conversations with G.~Lambiase and G.~F.~Aldi.
\section*{References}

\end{document}